\documentclass{Interspeech2024}

\interspeechcameraready

\title{\texorpdfstring{DualSpeech: Enhancing Speaker-Fidelity and Text-Intelligibility\\Through Dual Classifier-Free Guidance}{DualSpeech: Enhancing Speaker-Fidelity and Text-Intelligibility Through Dual Classifier-Free Guidance}}

\name[affiliation={1 *}]{Jinhyeok}{Yang}
\name[affiliation={1 *}]{Junhyeok}{Lee}
\name[affiliation={2 \dagger}]{Hyeong-Seok}{Choi}
\name[affiliation={1}]{Seunghun}{Ji}
\name[affiliation={1}]{Hyeongju}{Kim}
\name[affiliation={1}]{Juheon}{Lee}

\address{
  $^1$Supertone Inc., Republic of Korea\\
  $^2$ElevenLabs Inc., USA
} 
\email{yangyangii@supertone.ai, jun.hyeok@supertone.ai}

\keywords{text-to-speech, diffusion, classifier-free guidance}

\makeatletter
\DeclareRobustCommand\onedot{\futurelet\@let@token\@onedot}
\def\@onedot{\ifx\@let@token.\else.\null\fi\xspace}

\def\etal{\emph{et al}\onedot}
\makeatother

\begin{document}

\maketitle

\begin{abstract}
    Text-to-Speech (TTS) models have advanced significantly, aiming to accurately replicate human speech's diversity, including unique speaker identities and linguistic nuances. Despite these advancements, achieving an optimal balance between speaker-fidelity and text-intelligibility remains a challenge, particularly when diverse control demands are considered. Addressing this, we introduce DualSpeech, a TTS model that integrates phoneme-level latent diffusion with dual classifier-free guidance. This approach enables exceptional control over speaker-fidelity and text-intelligibility. Experimental results demonstrate that by utilizing the sophisticated control, DualSpeech surpasses existing state-of-the-art TTS models in performance. Demos are available at \url{https://bit.ly/48Ewoib}.
\end{abstract}

\def\thefootnote{*}\footnotetext{Equal contribution \, \, \textdagger Work done at Supertone Inc.}
\def\thefootnote{}\footnotetext{This research was supported by Culture, Sports and Tourism R\&D Program through the Korea Creative Content Agency grant funded by the Ministry of Culture, Sports and Tourism in 2022 (Project Name: Cultural Technology Research and Development, Project Number: R2022020066, Contribution Rate: 50\%)}
\def\thefootnote{\arabic{footnote}}

\vspace{-1.5pt}
\section{Introduction}
\vspace{-1.5pt}

Human speech is characterized by a wide array of variations, including distinctive speaker identities, different speech rhythms, tones, languages, and more. The goal of Text-to-Speech (TTS) is to emulate this richness, synthesizing speech that is not only natural and human-like but also encompasses a broad spectrum of these qualities and nuances.

Accordingly, state-of-the-art TTS models should excel in producing speech that not only captures the essence of the speaker, including their timbre, speaking style, accent, and emotions, for high speaker-fidelity but also ensures that the speech is easily understood, maintaining strong text-intelligibility \cite{valle,voicebox,clamtts,speartts,hierspeechpp, styletts2}. However, achieving a perfect balance between speaker-fidelity and text-intelligibility can be challenging in some cases. For example, using a yawning young woman's recording as a reference for speech synthesis might lead us into a dilemma: focusing too much on matching her voice, including the yawn, might compromise the clarity of the speech (text-intelligibility). Conversely, concentrating on making the speech clear could result in losing the yawn's unique effect, thereby producing speech that accurately represents a young woman's voice but lacks the intended distinctive characteristic, thus affecting speaker-fidelity.

Most TTS research to date evaluates these two components—speaker-fidelity and text-intelligibility—using metrics such as speaker similarity, naturalness MOS, and word error rate (WER); however, there has been little exploration into methods for independently controlling each element when they come into conflict. We believe that the ability to independently manipulate these factors would be highly beneficial in real-world TTS scenarios. To address this, we have sought out methodologies that enable such control, focusing on: 1) the use of representation disentanglement within generative models to separate and independently manage different aspects of speech, and 2) the application of classifier-free guidance in diffusion-based generative models, which allows for the independent conditioning and control of various conditions, actively exploring these approaches for practical solutions.

First, NANSY\footnote{For clarity, all NANSY referred to in this paper is NANSY++ \cite{nansypp}, rather than its earlier version \cite{nansy}.
} \cite{nansypp, nansy} has demonstrated quality improvements through self-supervised reconstruction from disentangled features.
Notably, NANSY stands out for providing high controllability via interpretable features such as linguistic features, fundamental frequency (f0), periodic and aperiodic amplitudes, and timbre features.
In particular, NANSY-TTS, an application of NANSY for TTS tasks, exemplifies NANSY's ability to independently manage timbre features, allowing for the disentanglement of speaking style and timbre. This capability affords enhanced controllability over the representation of various speakers. However, similar to broader challenges in the field, NANSY-TTS still grapples with controlling the balance between speaker-fidelity and text-intelligibility.

Second, diffusion models \cite{ddpm} employ classifier-free guidance (CFG) \cite{cfg} to control various conditions, a technique also adopted by speech diffusion models \cite{voicebox,voiceldm} for enhanced condition manipulation.
Specifically, VoiceLDM \cite{voiceldm} introduces dual CFG mechanism that allows separate control of environmental and content conditions. This feature allows VoiceLDM to manipulate the intensity of environmental and content conditions independently.
While our approach is similar to dual CFG of VoiceLDM, our method is more ideal in TTS by enabling control between text-intelligibility that follows text content and speaker-fidelity of reference speech.

In this paper, we introduce DualSpeech, a latent diffusion-based TTS model that achieves enhanced speaker-fidelity and text-intelligibility by utilizing dual classifier-free guidance. To attain high controllability with dual CFG, we introduce two phoneme-level conditioners; reference conditioner and text conditioner.
These networks are designed to model prior latent highly dependent on reference speech and text, respectively.
Through these networks, at the inference stage, we can manipulate the prosody of the generated speech to follow either the reference or the content by selecting CFG weights.
Our proposed approach demonstrates superior zero-shot TTS capability, along with enhanced intelligibility and controllability.

\vspace{-1.5pt}
\section{Method}
\vspace{-1.5pt}

\begin{figure*}[ht!]
  \centering
  \includegraphics[width=0.90\textwidth]{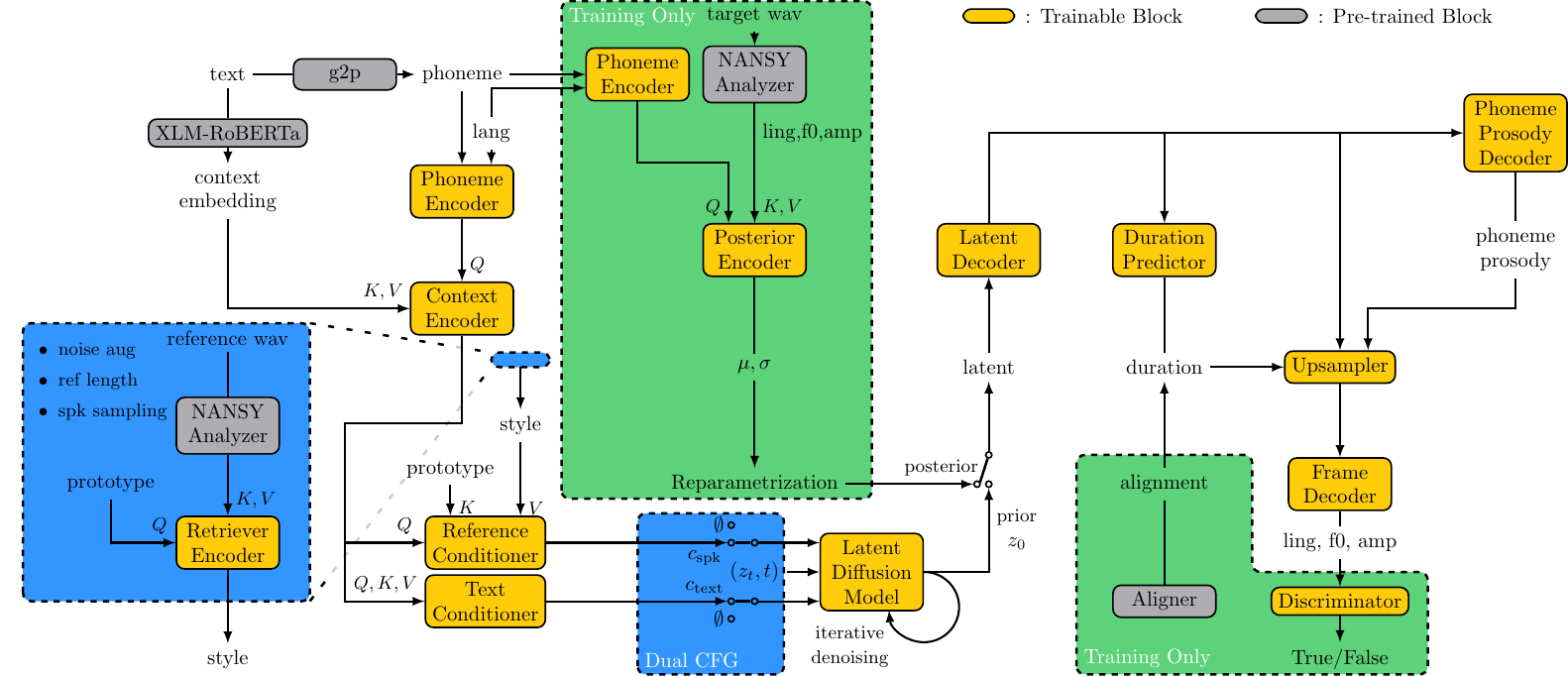}
    \vspace{-0.8\baselineskip}
  \caption{Overall model architecture of DualSpeech. Trainable blocks are colored in yellow and pre-trained modules are colored in gray. All blocks are based on the Transformer encoder architecture, even if their architecture is not mentioned in the main text.}
  \label{fig:model}
  
\vspace{-1.1\baselineskip}
\end{figure*}

DualSpeech is composed of three main components: NANSY, variational auto-encoder (VAE) \cite{vae}, and latent diffusion model (LDM) \cite{ldm}. 
The comprehensive architecture is illustrated in Figure \ref{fig:model}.
Unless specified otherwise, this paper assumes that the referenced modules are Transformer encoders, for which architectural details have not been provided.

Different from almost of TTS model, which generates mel-spectrogram, DualSpeech utilizes NANSY features.
We leverage a pre-trained NANSY for extracting NANSY features, aligning with the NANSY-TTS, which generates linguistic feature, f0, periodic amplitude, and aperiodic amplitude \cite{nansypp}.
In addition, Aligner is trained to align NANSY linguistic features with phonemes using monotonic alignment search (MAS) \cite{glowtts}.
Building on these pre-trained models, our model is trained in two stages, VAE training, and LDM training.
Our VAE, featuring a phoneme-level bottleneck, reconstructs NANSY features from given speech and phonemes.
Lastly, LDM generates VAE latent from given transcription and reference speech.

\vspace{-1.5pt}
\subsection{Phoneme-Level Variational Auto-Encoder} \label{section:vae}
\vspace{-1.5pt}

The VAE in DualSpeech processes inputs comprising an IPA sequence, which has been converted from text, and the corresponding NANSY features of the speech, to reconstruct the NANSY features of that speech.
Our VAE utilizes a phoneme-level bottleneck, inspired by \cite{finegrainedvae,diffvoice}.
This bottleneck is implemented through the cross-attention mechanism of the Transformer encoder, which uses the output of the phoneme encoder as a query and concatenated NANSY features as both key and value. 
This approach offers two advantages over frame-level models.
Firstly, a phoneme-based representation reliably conveys semantic information, as phonemes are symbolic representations of speech sounds. 
Secondly, it provides computational efficiency compared to models that learn at the frame-level,
as the computation complexity of the Transformer encoder scales as $\mathcal{O}(L^2)$ \cite{transformer}, where $L$ is the sequence length.
From the phoneme-level bottleneck, the posterior latent is sampled by estimated mean and variance.

The VAE decoder comprises a latent decoder, a duration predictor, a phoneme prosody decoder, an upsampler, and a frame decoder.
The decoding process begins with the latent variables passing through the latent decoder.
The outputs from this network feed into the duration predictor, phoneme prosody decoder, and upsampler in parallel.
The duration predictor and phoneme prosody decoder are responsible for estimating duration and f0 at the phoneme-level, respectively.
At the upsampler stage, phoneme-level output of the latent decoder is upsampled to a frame-level sequence by ground truth duration from the pre-trained MAS aligner. 
The architecture of the upsampler is almost identical to the learned upsampler from Parallel Tacotron 2 \cite{paralleltacotron2}, excluding the channel dimension.
The upsampled frame-level feature is then input into the frame decoder to reconstruct NANSY features. 

In addition, we enhance the performance of the VAE through adversarial training \cite{gan, himuv}.
Our discriminator consists of a simple convolution network trained using least-square loss and feature-matching loss.

Our VAE model is trained with NANSY feature reconstruction losses, phoneme-level f0 reconstruction loss, duration loss, KL divergence of latent, and adversarial losses.

\vspace{-1.5pt}
\subsection{Phoneme-Level Latent Diffusion Model}
\vspace{-1.5pt}

DualSpeech's LDM is trained to estimate the phoneme-level posterior latent generated by pre-trained VAE discussed in Section \ref{section:vae}. 
Also in LDM, the phoneme-level model significantly reduces computational demand compared to frame-level models by decreasing the computation required for iterative denoising, which is a major bottleneck of diffusion models.
To simultaneously achieve naturalness and speaker similarity once by generating prior latent through LDM, our model is structured into two main components: conditioners and conditional diffusion model with dual CFG. 

Conditioners include two types: the reference conditioner and the text conditioner.
To inject conditional information for both the reference speaker and the text, these conditioners are designed to produce phoneme-wise conditions.
These conditioners share inputs from the context encoder, which is a Transformer encoder employing cross-attention to model context-aware features derived from the outputs of the phoneme encoder and context embeddings. 
For obtaining context embeddings, we utilize pre-trained XLM-RoBERTa \cite{xlmroberta}.

Given that the text conditioner relies solely on text inputs, we expect that by adjusting $\omega_\mathrm{text}$, the CFG weight for the text conditioner's output $c_\mathrm{text}$, we can control fine intelligibility.
At the reference conditioner, we aim to generate speaker-aware phoneme-wise conditioning to facilitate zero-shot capability.
To capture speaker's style from reference speech and enable zero-shot capabilities, we integrate Retriever \cite{retriever} into our reference conditioner.
Reference speech is sampled from the target speaker's subset, noise corrupted, and cut into random lengths to reduce training-inference mismatch.
We extract NANSY features from the reference speech and then feed them into the cross-attention mechanism of the retriever encoder.
The query for this cross-attention is fixed-length tokens, referred to as prototypes, which in our case is 60.
Consequently, the output of this Transformer is also fixed-length tokens encapsulating the reference speech's speaker style.
Moreover, the reference conditioner encodes speaker-related conditions by leveraging these speaker tokens as a value for cross-attention and employs prototypes of identical length to those used by the retriever encoder. 
Analogous to $\omega_\mathrm{text}$ in the text conditioner, the similarity to the speaker can be modulated by adjusting $\omega_\mathrm{spk}$, the CFG weight for reference conditioner's output $c_\mathrm{spk}$.

Our diffusion model's architecture is also based on a Transformer encoder, akin to that of DiT \cite{dit}.
Furthermore, instead of the conditioning mechanism from DiT's adaptive layer norm, we change it to a simpler addition of conditions after two MLP layers, similar to that of DiffWave \cite{diffwave}.

Our LDM is trained using the $L_1$ loss as WaveGrad \cite{wavegrad}:
\begin{equation}
    \mathcal{L} = \left\lVert \epsilon-\epsilon_{\theta}\left(\sqrt{\bar \alpha_t} \mu + \sqrt{1- \bar \alpha_t} \epsilon ,t, c_{\mathrm{spk}}, c_{\mathrm{text}}\right)\right\rVert_1,
\end{equation}
where $\epsilon \sim \mathcal{N}(0,I)$ is the added noise, $\epsilon_\theta$ is the diffusion model with parameters, $\mu$ is the mean estimated by the VAE, $t$ denotes the timestep, and $\bar \alpha_t$ corresponds to the noise coefficient at time $t$.
We implement random dropout for both $c_{\mathrm{text}}$ and $c_{\mathrm{spk}}$ to employ CFG during the inference.
Specifically, we drop $c_{\mathrm{text}}$ by 5\% and $c_{\mathrm{spk}}$ by 10\%, with an additional 10\% dropout applied to both to promote the frequency of null-conditioned scenarios.
Training employs a discrete integer diffusion timestep and a noise schedule.
$t$ is uniformly sampled from $[1, T]$, where $T=200$.
Following the approach of prior diffusion models \cite{ddpm, diffwave,diffwave_fast}, we adopt a linear variance schedule defined as $\beta_i = \beta_1 + (\beta_T-\beta_1)(i-1)/(T-1)$, setting $\beta_1=0.0001$ and $\beta_T = 0.03$.
The noise coefficient $\bar\alpha_t$ is calculated as $\bar\alpha_t=\prod_{i=1}^t \alpha_i = \prod_{i=1}^t (1-\beta_i)$.

\vspace{-1.5pt}
\subsection{Dual Classifier-Free Guidance for TTS}
\vspace{-1.5pt}
Our method extends to generate latent with fine control between text and reference conditions.
Inspired by Lee \etal \cite{voiceldm}, DualSpeech employs a dual CFG for TTS, which can be represented as follows:
\begin{align}
    \Tilde{\epsilon}_{\theta}(&z_t,t, c_{\mathrm{spk}}, c_{\mathrm{text}}) = \epsilon_{\theta}(z_t,t, c_{\mathrm{spk}}, c_{\mathrm{text}}) \nonumber \\
    &+ \omega_{\mathrm{spk}} \bigl( \epsilon_{\theta}(z_t,t, c_{\mathrm{spk}}, \emptyset) - \epsilon_{\theta}(z_t, t, \emptyset, \emptyset) \bigr) \nonumber \\
    &+ \omega_{\mathrm{text}} \bigl( \epsilon_{\theta}\bigl(z_t,t, \emptyset, c_{\mathrm{text}}\bigr) - \epsilon_{\theta} \bigl( z_t,t, \emptyset, \emptyset\bigr) \bigr),
\end{align}
where $\Tilde{\epsilon}_{\theta}$ represents the classifier free-guided noise,  $z_t$ is a latent at timestep $t$ defined as $z_t = \sqrt{\bar \alpha_t} \mu + \sqrt{1- \bar \alpha_t} \epsilon$, and $\emptyset$ denotes zero tensors corresponding to a null-conditioned state.
A noticeable difference from VoiceLDM lies in its conditioning from descriptions of the acoustic environment instead of the speaker's style itself.
This constraint is influenced by CLAP \cite{clap}, which is trained not only with captions describing the speaker's style but on a broad range of audio files and their captions. This includes the undesired noisy acoustic environments that are generally adverse to the objectives of TTS. 
In contrast to VoiceLDM, our approach allows for a more granular manipulation of speech synthesis, directly addressing the challenge of balancing text and speaker similarity.

\vspace{-1.5pt}
\subsection{Inference}
\vspace{-1.5pt}

During inference, only the LDM, the VAE decoder, and the NANSY synthesizer are utilized. 
The LDM generates phoneme-level latent through iterative denoising.
We employ fast sampling suggested by Kong and Ping \cite{diffwave_fast}, utilizing a variance noise schedule as [1e-4, 5e-4, 1e-3, 5e-3, 0.01, 0.02, 0.05, 0.2, 0.3, 0.5, 0.4, 0.3, 0.3, 0.2, 0.1, 0.1].
The generated prior latent is then processed by the VAE decoder, which includes upsampling to the NANSY frame-level, and estimation of linguistic, f0, and amplitudes of NANSY.
Finally, a raw waveform is synthesized by the pre-trained NANSY synthesizer.

\vspace{-1.5pt}
\section{Experiments}
\vspace{-1.5pt}
\subsection{Settings}
\vspace{-1.5pt}
\subsubsection{Training}
\vspace{-1.5pt}
All of our experiments were conducted on 8 NVIDIA RTX 4090 GPUs, utilizing dynamic batch sizes throughout the training process.
Our pre-trained NANSY is trained with an identical setup following Choi \etal \cite{nansypp}.
Our model processes input at a sampling rate of 16 kHz and generates outputs at a sampling rate of 44.1 kHz.
The fundamental frequency is converted to the MIDI scale and then divided by 84, corresponding to 1,046 Hz.
We applied an internal grapheme-to-phoneme (G2P) model to convert grapheme text to an IPA-based phoneme sequence.

\vspace{-1.5pt}
\subsubsection{Dataset}
\vspace{-1.5pt}
For the training of our TTS model, we utilized four datasets: LJSpeech \cite{ljspeech}, VCTK \cite{vctk}, Hi-Fi TTS \cite{hifitts}, and LibriTTS \cite{libritts}. These datasets encompass a wide range of speaker characteristics, including pronunciation, accents, timbre, and prosody, as well as linguistic nuances, offering a comprehensive diversity of English speech.
The total dataset consists of 945 hours of high-quality speech and 2,576 English speakers, including a broad spectrum of English accents.

To evaluate how much the proposed method improves naturalness and similarity, a carefully curated set of 9 speakers exhibiting a broad spectrum of vocal characteristics was selected. The speaker set includes unseen non-English speakers and speakers with emotional tones such as sleepy.

Following previous studies \cite{valle,voicebox,clamtts,speartts}, we utilize a subset of the LibriSpeech test-clean dataset for objective evaluation.
This subset contains speech clips with durations ranging from 4 to 10 seconds.

\vspace{-1.5pt}
\subsubsection{Evaluation Metrics}
\vspace{-1.5pt}
We assess the controllability of speaker similarity and naturalness through three distinct mean opinion scores (MOS): quality MOS (QMOS), which assesses sound quality, speaker similarity MOS (SMOS), which evaluates the similarity between the speaker of the prompt and the generated speech, and prosody MOS (PMOS), which gauges the naturalness of the speech's prosody. To ensure a fair evaluation of audio with various sampling rates, all audio samples were downsampled to 16 kHz before being assessed. For evaluation, the LibriTTS test-other subset was utilized as the input text. 
To assess the ground truth (GT) for QMOS, reference speeches that were used as speaker prompts were evaluated. Similarly, for PMOS GT evaluation, LibriTTS speech samples corresponding to the input text were measured.

To assess the correctness and intelligibility of the generated speech, we measure the word error rate (WER) and character error rate (CER) by comparing the transcribed text of the generated speech with the corresponding input text.
We transcribe speech by pre-trained CTC-based HuBERT-Large\footnote{https://huggingface.co/facebook/hubert-large-ls960-ft} \cite{hubert}.

\vspace{-1.5pt}
\subsection{Results}
\vspace{-1.5pt}
\subsubsection{Subjective Evaluation}
\vspace{-1.5pt}
We conducted three MOS tests for subjective evaluation, comparing our model against models with official implementations, including those that are state-of-the-art models \cite{hierspeechpp,styletts2,yourtts}. 
The distinctive feature of DualSpeech lies in its capability to precisely modulate the balance between text content and speaker characteristics with the CFG weights ($\omega_\mathrm{text}$ and $\omega_\mathrm{spk}$), thereby enabling synthesis to adapt to diverse application scenarios.

Our system consistently maintains a high level of quality in terms of QMOS, while also demonstrating the ability to selectively enhance either PMOS or SMOS through strategic weight adjustments. 
In configurations prioritizing text content with $(\omega_\mathrm{text},\omega_\mathrm{spk})=(4.0,1.0)$, our system not only achieves a QMOS of 4.24, indicative of superior sound quality but also achieves a PMOS of 3.83, underscoring its exceptional proficiency in replicating natural prosody.
This suggests our system's adeptness at capturing and reproducing the nuanced tones and rhythms inherent in phonemes.
Moreover, it enables us to faithfully replicate the timer of speakers while excluding biased expressions found in reference speech, such as yawning, and instead generate neutral expressions derived from the training datasets with the reference's timbre.

Conversely, when emphasizing speaker characteristics with $(\omega_\mathrm{text},\omega_\mathrm{spk})=(1.0,4.0)$, while maintaining the same level of QMOS, our system significantly improves SMOS to 3.74.
This underscores its outstanding ability to capture speaker similarity.
The noticeable enhancement in SMOS accentuates the system's capability to replicate distinct voice traits of speakers, which is vital for personalized voice synthesis applications.

\begin{table}[t]
\caption{Subjective results for zero-shot TTS}
\vspace{-4.8pt}
\centering
 \resizebox{0.95\linewidth}{!}{%
\begin{tabular}{l|c|c|c|c|c}
\toprule
& $\omega_\mathrm{text}$ & $\omega_\mathrm{spk}$ & QMOS & SMOS & PMOS \\
\midrule
GT              &-&-     & 3.92$\scriptstyle{\pm0.15}$    & - & 4.34$\scriptstyle{\pm0.13}$ \\
\midrule
YourTTS         &-  &-   & 2.41$\scriptstyle{\pm0.13}$  & 1.40$\scriptstyle{\pm0.12}$ & 1.98$\scriptstyle{\pm0.14}$ \\
HierSpeech++    &-  &-   & 3.62$\scriptstyle{\pm0.12}$  & 3.31$\scriptstyle{\pm0.19}$ & 3.01$\scriptstyle{\pm0.16}$ \\
StyleTTS 2      &-  &-   & 3.95$\scriptstyle{\pm0.12}$  & 1.84$\scriptstyle{\pm0.17}$ & 3.76$\scriptstyle{\pm0.14}$ \\
DualSpeech (Ours)            &1.0&4.0 & 4.18$\scriptstyle{\pm0.11}$  & \bf{3.74}$\scriptstyle{\pm0.21}$ & 3.36$\scriptstyle{\pm0.17}$ \\ 
DualSpeech (Ours)            &4.0&1.0 & \bf{4.24}$\scriptstyle{\pm0.12}$  & 2.35$\scriptstyle{\pm0.22}$ & \bf{3.83}$\scriptstyle{\pm0.15}$ \\
\bottomrule
\end{tabular}
}
\label{subjective}
\vspace{-4.8pt}

\end{table}

\vspace{-1.5pt}
\subsubsection{Objective Result}
\vspace{-1.5pt}
In TTS studies, objective measures like WER and CER serve as critical benchmarks for evaluating the robustness and correctness of synthesized speech.
Our study presents an extensive objective evaluation conducted on a subset of the LibriSpeech dataset, the results of which are detailed in Table \ref{tab:objective}.
This evaluation underscores the efficacy of our proposed TTS system, particularly when compared against state-of-the-art systems \cite{valle,voicebox,clamtts,speartts,yourtts} and GT.

The GT recordings exhibit low WER and CER, at 2.26 and 0.61 respectively, setting a high standard for speech synthesis quality. Among the competing systems, Voicebox achieves an impressive WER of 2.00 in one instance, the lowest among the synthesized voices, albeit without a corresponding CER reported.
Our system, under various configurations of text and speaker CFG weights, demonstrates competitive performance, particularly with a configuration of $(\omega_\mathrm{text},\omega_\mathrm{spk})=(2.0,1.0)$, achieving a WER of 2.59 and a CER of 0.77. These results are notably close to the GT, highlighting our system's ability to maintain high levels of speech intelligibility and accuracy.

Furthermore, our system's adaptability is evident in its performance across different configurations, suggesting that precisely adjusting the balance between text and speaker emphasis can optimize performance for specific applications. While no single configuration universally outperforms all others, the ability to adjust these parameters allows for significant flexibility in tailoring the system to meet diverse needs.

\begin{table}[t]
  \caption{Evaluating LibriSpeech-subset for robustness and accuracy with highest scores in bold, second highest underlined, and baseline scores asterisked.}
  \vspace{-4.8pt}
\centering
\resizebox{0.95\linewidth}{!}{%
\begin{tabular}{l|c|c|c|c}
\toprule
& $\omega_\mathrm{text}$ & $\omega_\mathrm{spk}$ & WER$\downarrow$ & CER$\downarrow$  \\
\midrule
GT         &-&-     & \textbf{2.26}  & \textbf{0.61}    \\
\midrule

YourTTS$^*$ \cite{yourtts,clamtts}  &-&- & 7.92  & 3.18    \\ 
VALL-E$^*$ \cite{valle}  &-&- & 5.9\phantom{0}  & -    \\ 
Voicebox$^*$ \cite{voicebox}   &-&- & \textbf{1.90}  & -    \\ 
CLaM-en$^*$ \cite{clamtts}  &-&- & 5.11  & 2.87    \\ 
SPEAR-TTS$^*$ \cite{speartts}   &-&- & -  & 1.92    \\ 
DualSpeech (Ours) &1.0&1.0   & 2.77  & 0.83          \\ 
DualSpeech (Ours) &1.0&2.0  & 2.62  & 0.81         \\
DualSpeech (Ours) &2.0&1.0  & \underline{2.59}  & \textbf{0.77}   \\
DualSpeech (Ours) &2.0&2.0  & 2.62  & \underline{0.80}  \\
\bottomrule
\end{tabular}
}
\label{tab:objective}
\vspace{-9.5pt}
\end{table}

In addition to speech synthesis quality, inference speed is a crucial factor for the practical application of TTS systems. Table \ref{inferencespeed} shows our phoneme-level diffusion model's superior inference speed, clocking in at 0.19 seconds, significantly faster than other frame-based diffusion models or auto-regressive language models, including Voicebox and CLaM-en, which report inference times of 6.4 and 4.2 seconds, respectively. This remarkable speed does not compromise the quality of the synthesized speech, positioning our system as a highly efficient and effective solution for real-time TTS applications.

\begin{table}[t]
\caption{Inference speed of models. CLaM-en and DualSpeech were tested on an A100, and Voicebox's GPU details are undisclosed.}
\vspace{-4.8pt}
\resizebox{0.95\linewidth}{!}{%
\centering

\begin{tabular}{l|c c c }
\toprule
& Voicebox$^*$ \cite{voicebox}  & CLaM-en$^*$ \cite{clamtts} &  DualSpeech (Ours) \\
\midrule
Inference Time (s)  & 6.4 (64 NFE) & 4.2 & \textbf{0.19} (16 steps)\\
\bottomrule
\end{tabular}
}
\label{inferencespeed}

\vspace{-4.8pt}
\end{table}
\vspace{-1.5pt}
\section{Conclusion}
\vspace{-1.5pt}
In this work, we introduce DualSpeech, a text-to-speech model that combines a phoneme-level latent diffusion model with dual classifier-free guidance (CFG). This model showcases exceptional zero-shot TTS capabilities, excelling in speaker-fidelity and text-intelligibility. DualSpeech provides high-quality voice synthesis with the flexibility to adjust for either speaker-fidelity or text-intelligibility, according to specific requirements. We believe that integrating our dual CFG approach into any diffusion-based TTS system will significantly refine the balance between speaker fidelity and text intelligibility.

\clearpage



\bibliographystyle{IEEEtran}

\end{document}